# Magnetic and transport properties of *n*-type Fe-doped In$_2$O$_3$ ferromagnetic thin films

Xiao-Hong Xu[1*], Feng-Xian Jiang[1,2], Jun Zhang[1], Xiao-Chen Fan[1], Hai-Shun Wu[1], G. A. Gehring[3]

1. School of Chemistry and Materials Science, Shanxi Normal University, Linfen 041004, P. R. China
2. College of Chemistry and Chemical Engineering, Taiyuan University of Technology, Taiyuan 030024, P. R. China.
3. Department of Physics and Astronomy, University of Sheffield, Hicks Building, Sheffield S3 7RH, UK

Room temperature ferromagnetism was observed in *n*-type Fe-doped In$_2$O$_3$ thin films deposited on *c*-cut sapphire substrates by pulsed laser deposition. Structure, magnetism, composition, and transport studies indicated that Fe occupied the In sites of the In$_2$O$_3$ lattice rather than formed any metallic Fe or other magnetic impurity phases. Magnetic moments of films were proved to be intrinsic and showed to have a strong dependence on the carrier densities which depended on the Fe concentration and its valance state as well as oxygen pressure.

Dilute magnetic semiconductors (DMSs) have attracted wide interest due to their potential applications in spintronics.[1] Since room temperature ferromagnetism in Mn doped ZnO had been theoretically predicted by Dietl *et al.*,[2] a variety of transition metals (TM) doped oxides, such as ZnO,[3,4] TiO$_2$,[5] and SnO$_2$[6] were extensively investigated. However, due to the small solubility of TM in these host semiconductors, the origin of ferromagnetism might be attributed to magnetic impurities.[7,8] Therefore, DMSs based on host semiconductors with high solubility of TM ions are highly desirable. Fortunately, the concentration of Fe ions in the In$_2$O$_3$ host lattice was found to be as high as 20%.[9] However, the magnetic properties of Fe-doped In$_2$O$_3$ reported by different research groups are quite controversial. Xing *et al.*[10] observed high temperature ferromagnetism in Fe-doped In$_2$O$_3$ films and Jayakumar *et al.*[11] found room temperature ferromagnetism in Fe-doped In$_2$O$_3$ powders. Whereas, Peleckis *et al.*[12] found only paramagnetism in Fe-doped In$_2$O$_3$ bulk and Kohiki *et al.*[13] attributed the ferromagnetism to the presence of ferromagnetic Fe$_2$O$_3$ nanoclusters in In$_2$O$_3$ matrix. These inconsistent results are likely to be due to the fact that the magnetic properties of Fe-doped In$_2$O$_3$ are extremely sensitive to the fabrication methods or experimental conditions. In addition, the origin of ferromagnetism in this system is still under debate.

We have recently reported that Fe-doped In$_2$O$_3$ powders are paramagnetic when sintered in air and become magnetic when annealed in vacuum.[14] It is strongly believed that the oxygen vacancies in the samples play a critical role in inducing ferromagnetism. In this letter, In$_2$O$_3$ films with different Fe-doped concentrations were fabricated by a pulsed laser deposition (PLD) under various oxygen partial pressures ($P_{O_2}$). In Fe-doped In$_2$O$_3$ films deposited by PLD, defects may from several sources, such as oxygen vacancies, Fe$^{3+}$, reduced Fe$^{2+}$, and interstitial In$^{3+}$. Therefore, it is essential to study systematically the effects of Fe concentration and $P_{O_2}$ on the structural, magnetic, composition and transport properties of (In$_{1-x}$Fe$_x$)$_2$O$_3$ films. In order to discuss the magnetic origin and mechanism, we focus on the investigation of the relationship between the transport and magnetization.

(In$_{1-x}$Fe$_x$)$_2$O$_3$ films were deposited on Al$_2$O$_3$ (0001) substrates by a PLD technique. The (In$_{1-x}$Fe$_x$)$_2$O$_3$ (*x*=0.02, 0.05, 0.1, 0.2) ceramic targets sintered at 1100 °C were ablated using a KrF excimer laser operating. During deposition, the substrate temperature was maintained at 600 °C and the $P_{O_2}$ was varied from 5×

[*]Author to whom correspondence should be addressed:
Electronic mail: xuxh@dns.sxnu.edu.cn

$10^{-3}$ to 100 mTorr. The structure of the films was analyzed by x-ray diffraction (XRD) with Cu Kα radiation (λ=0.15406 nm). The composition of the films was determined by x-ray photoelectron spectroscopy (XPS). The magnetic measurements were performed at room temperature using a vibrating sample magnetometer (VSM). The electrical properties of the films were determined by a Hall effect measurement system in the van der Pauw four-point configuration.

The XRD patterns of $(In_{1-x}Fe_x)_2O_3$ films deposited at a $P_{O_2}$ of 5×$10^{-3}$ mTorr are shown in Fig. 1(a). The single-phase cubic $In_2O_3$ structures with well-oriented (222) texture are observed, suggesting that Fe-doped $In_2O_3$ films retain the same crystalline structure as the un-doped $In_2O_3$ film. There are no other peaks related to Fe metal clusters or Fe oxide secondary phases in any of the films, even though the Fe concentration ($x$) is as high as 20%. A clear shift of (222) peaks to higher angles is observed, as shown in the inset of Fig. 1(a), which depends on the $x$. This reduction in the lattice constant with Fe doping is expected as the smaller Fe ions ($Fe^{3+}$ and $Fe^{2+}$ ionic radii are 0.079 and 0.092 nm, respectively, $In^{3+}$ is 0.094 nm) are incorporated into the In sites of the $In_2O_3$ lattice.

Figure 1(b) shows the XRD patterns of 2% Fe-doped $In_2O_3$ films grown at different $P_{O_2}$. In its inset, a section of XRD with 2θ ranging from 29 to 32° is also plotted. It is interesting that the (222) peaks move to lower angles with increasing $P_{O_2}$ from 5×$10^{-3}$ to 10 mTorr, and then shift to higher angles as $P_{O_2}$ is increased to 100 mTorr. This means that the d $_{(222)}$ values increase first and then drop down at a critical point of 10 mTorr; this phenomenon also occurs for other Fe concentrations samples.

XPS studies are used to determine the composition and oxidation states of the iron in the films. At the lowest $P_{O_2}$ of 5×$10^{-3}$ mTorr, the binding energy of Fe 2p$^{3/2}$ is of 710.4 eV that is between $Fe^{2+}$ (709.9 eV) and $Fe^{3+}$ (711.0 eV), which corresponds to a mixed oxidation state of +2 and +3.[15] However, at the highest $P_{O_2}$ of 100 mTorr, the binding energy of Fe 2p$^{3/2}$ is close to 711.0 eV, indicating that the iron has converted to $Fe^{3+}$.

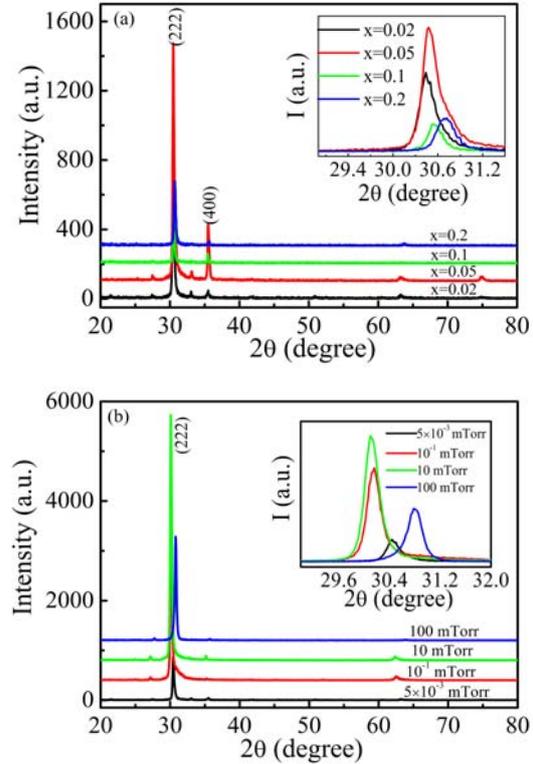

FIG.1. XRD patterns for $(In_{1-x}Fe_x)_2O_3$ films deposited at a $P_{O_2}$ of 5×$10^{-3}$ mTorr (a) and $(In_{0.98}Fe_{0.02})_2O_3$ films grown under different $P_{O_2}$ conditions (b).

An increase of $P_{O_2}$ during film growth leads to a reduction in the number of the oxygen vacancies together with an oxidation of $Fe^{2+}$ ions to $Fe^{3+}$ ions (namely the increase of the ratio of $Fe^{3+}/Fe^{2+}$). The reduction of the concentration of the oxygen vacancies leads to an increase of d $_{(222)}$ values, whereas the increase of $Fe^{3+}/Fe^{2+}$ ratio gives an opposite effect. Thus the net change in the lattice parameter is due to a competition between these two factors. The value of d $_{(222)}$ is small at low values of $P_{O_2}$ due to the presence of oxygen vacancies and the value of d $_{(222)}$ increases with oxygen pressure as the number of oxygen vacancies is reduced. However when the $P_{O_2}$ value is increased beyond 10 mTorr, all the $Fe^{2+}$ is oxidized to $Fe^{3+}$ ions, which causes a strong reduction in $d_{(222)}$.

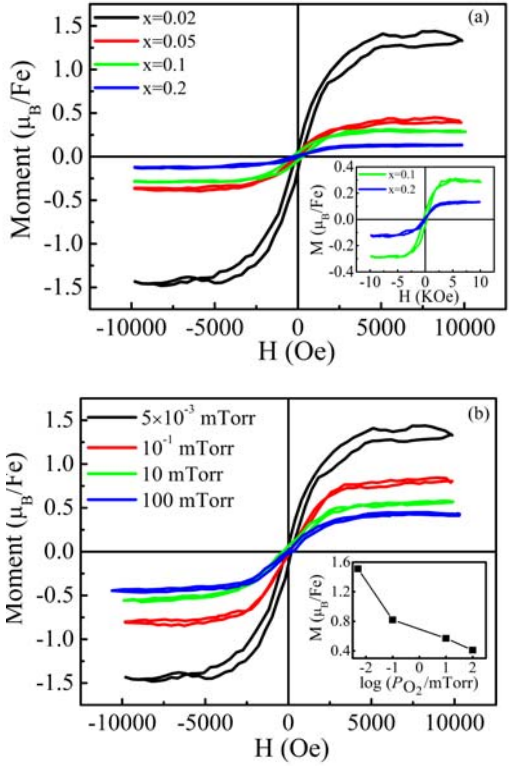

FIG. 2. Room temperature magnetization plots for $(In_{1-x}Fe_x)_2O_3$ films grown at $P_{O_2}$ of $5\times10^{-3}$ mTorr (a) and $(In_{0.98}Fe_{0.02})_2O_3$ films deposited at various $P_{O_2}$ (b).

Figure 2(a) depicts the room temperature hysteresis loops of the $(In_{1-x}Fe_x)_2O_3$ films grown under the $P_{O_2}$ of $5\times10^{-3}$ mTorr. The diamagnetic contribution from the substrate has been subtracted from the loops. The inset shows enlarged hysteresis loops of the 10 and 20% Fe-doped samples. The saturation magnetizations ($M_s$) of the 2, 5, 10 and 20% Fe-doped films are 1.51, 0.39, 0.32 and 0.14 $\mu_B$ per Fe atom, and the coercive fields ($H_c$) of them are 146, 81, 340 and 143 Oe, respectively. It reveals a strong reduction of $M_s$ with increasing Fe concentration, ruling out the extrinsic origin of the ferromagnetism from Fe clusters. The decrease in $M_s$ at higher Fe concentration could be attributed to an increase of the antiferromagnetic coupling between Fe ions, which has been predicted in earlier theoretical studies;[16] however another factor may be that the introduction of Fe reduces the carrier density as discussed later. The value of $M_s$ of 2% Fe film grown at $P_{O_2}=5\times10^{-3}$ mTorr is 1.51 $\mu_B$/Fe, which greatly exceeds the previous reported values of Fe doped $In_2O_3$ samples.[9-11] Figure 2(b) shows the magnetization versus applied magnetic field of the 2% Fe-doped $In_2O_3$ films deposited under different $P_{O_2}$. The variation of the $M_s$ with $P_{O_2}$ is shown in the inset of this figure, $M_s$ decreases from 1.51 to 0.41 $\mu_B$/Fe with the increase of $P_{O_2}$ from $5\times10^{-3}$ to 100 mTorr.

To gain an insight into the possible magnetic mechanism responsible for the observed ferromagnetic ordering in $(In_{1-x}Fe_x)_2O_3$ films, the electron carrier density ($n_c$) and resistivity ($\rho$) of $(In_{1-x}Fe_x)_2O_3$ films were measured at room temperature. Figure 3(a) shows the dependence of $\rho$ and $n_c$ on Fe concentration in $(In_{1-x}Fe_x)_2O_3$ films grown at $P_{O_2}$ of $5\times10^{-3}$ mTorr. As seen clearly from Fig. 3(a), when the Fe concentration is increased from 2 to 20%, $\rho$ increases and $n_c$ decreases monotonically. This might be explained by the presence of more Fe in +2 state in the high concentration samples, which act as deep level acceptors, and compensates the electrons arising from the oxygen vacancies in the films. In addition scattering from $Fe^{2+}$ may also contribute to the decrease of the mobility. Figure 3(b) shows the variation of $\rho$ and $n_c$ of $(In_{0.98}Fe_{0.02})_2O_3$ films deposited at various $P_{O_2}$; $n_c$ decreases and $\rho$ increases with an increase of $P_{O_2}$. The dependence of $\rho$ and $n_c$ on $P_{O_2}$ is easily understood that during film growth, $P_{O_2}$ directly determines the carrier density in the films.

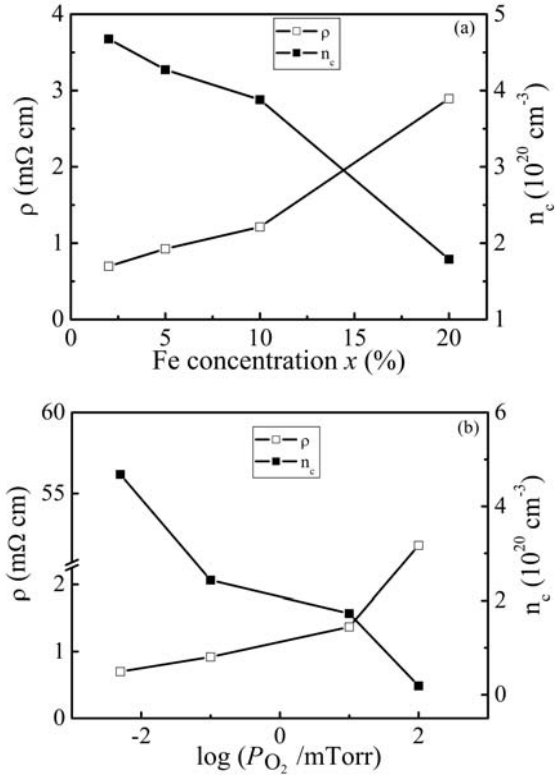

FIG. 3. The electrical resistivity, ρ, and carrier density, $n_c$, of $(In_{1-x}Fe_x)_2O_3$ films grown at $P_{O_2}$ of $5\times10^{-3}$ mTorr dependence of Fe concentration (a) and $(In_{0.98}Fe_{0.02})_2O_3$ films grown under various $P_{O_2}$ (b).

The metallic regime was identified by two tests: (i) the Fermi temperature $T_F$ should be high compared to room temperature, and (ii) the mean free path at room temperature, λ, estimated using the Drude formula $\lambda = \dfrac{\hbar k_F}{n_c e^2 \rho}$ (where $n_c$ and ρ are the measured carrier density and resistivity and $k_F$ is found from $n_c$), should be greater than the lattice spacing (3 Å).[3] We have estimated the $T_F$ and λ of our samples shown in Fig. 3. Clearly, all of them satisfy these two criteria, namely, λ>3 Å and $T_F$ is higher than 300 K.

Comparing the Fig. 2 with Fig. 3, we note that there is a strong relationship between $n_c$ and $M_s$; the value of $M_s$ decreases as $n_c$ is reduced. This confirms that the carrier concentration plays a significant role in the manifestation of ferromagnetism in Fe-doped $In_2O_3$ films and suggests that the carrier-induced mechanism is suitable for explaining the observed ferromagnetism in the metallic n-type $(In_{1-x}Fe_x)_2O_3$ films.

In summary, an extensive study was done on the structure, magnetism and composition and transport of $(In_{1-x}Fe_x)_2O_3$ films as a function of Fe concentration and oxygen partial pressure. We presented evidence to show that Fe ions are on the In sites of the $In_2O_3$ lattice and that the observed ferromagnetism is intrinsic rather than from the presence of metallic Fe clusters or Fe oxides secondary phases in the films. The dependence of the magnetization with carrier density supports a carrier-mediated model.

The authors would like to express their thanks to the NSFC (60776008), the NCET-07-0527, the RFDP (20070118001) and the NSFSX (2008011042-1).